# Navigating the Landscape of Distributed File Systems: Architectures, Implementations, and Considerations

Xueting Pan[1] and Ziqian Luo[1,*], Lisang Zhou[2]

[1]Oracle, Seattle, 98101, WA, United States, Email: xtpan8800@gmail.com
[1]Oracle, Seattle, 98101, WA, United States, Corresponding author, Email: luoziqian98@gmail.com
[2]Bazaarvoice Inc., Austin, 78759, TX, United States, Email: lzhou@berkeley.edu

**Abstract:** Distributed File Systems (DFS) have emerged as sophisticated solutions for efficient file storage and management across interconnected computer nodes. The main objective of DFS is to achieve flexible, scalable, and resilient file storage management by dispersing file data across multiple interconnected computer nodes, enabling users to seamlessly access and manipulate files distributed across diverse nodes. This article provides an overview of DFS, its architecture, classification methods, design considerations, challenges, and common implementations. Common DFS implementations discussed include NFS, AFS, GFS, HDFS, and CephFS, each tailored to specific use cases and design goals. Understanding the nuances of DFS architecture, classification, and design considerations is crucial for developing efficient, stable, and secure distributed file systems to meet diverse user and application needs in modern computing environments.

## 1. Introduction

A distributed file system (DFS)[1, 2] is a sophisticated solution crafted to efficiently store and manage file data across numerous interconnected computer nodes. By dispersing file data across multiple nodes connected via a network, DFS achieves the goal of decentralized file storage management. Its primary aim is to furnish an agile, scalable, and resilient file storage framework, empowering users to seamlessly access and manipulate files distributed across diverse nodes.

DFS orchestrates communication and coordination among networked nodes to facilitate distributed storage, access, and administration of files. These nodes encompass physical servers, virtual machines, or cloud computing instances. Typically, DFS employs a unified file system namespace, abstracting away the complexities of file storage locations, thus enabling users to interact with distributed file systems as effortlessly as with local file systems.

Within a DFS, files are often partitioned into discrete data blocks and distributed across assorted nodes to fortify data reliability and accessibility. DFS incorporates fault-tolerance mechanisms to detect and autonomously rectify node failures or data anomalies, thereby upholding data integrity and dependability. Furthermore, DFS integrates concurrency control mechanisms to enable multiple users to execute concurrent read and write operations on files.

In essence, DFS serves as a pivotal file storage and administration system in distributed computing environments, poised to deliver optimal performance, scalability, and reliability in addressing the burgeoning demands of contemporary large-scale data processing and storage requirements.

The structure of the article is as follows: the second section describes the key components of high-level distributed file system architecture, the third section focuses on discussing the classification of distributed file systems based on architecture. The fourth section discusses various factors to consider when designing distributed file systems, and the fifth section introduces several common distributed file systems.

Additionally, the application of deep learning models in DFS involves leveraging these models for data processing tasks such as data cleaning, preprocessing, and feature extraction [3,4,5]. Through techniques like noise removal, normalization, and feature extraction using models like CNNs and RNNs, deep learning enhances data quality and extracts valuable insights from files stored within DFS environments [6,7].

## 2. High-level distributed file system architectures

In high-level distributed file system architecture, several key components and design considerations are typically involved:

### 2.1. NameNode

The NameNode is the central node of the distributed file system, responsible for maintaining the file system's namespace and metadata information, including crucial attributes like file and directory names, locations, sizes, and permissions. It also receives file operation requests from clients, such as file creation, deletion, renaming, and locating, forwarding them to the appropriate DataNodes while managing file access permissions and ensuring metadata consistency.

### 2.2. DataNodes

Under the management of the NameNode, DataNodes store the actual data blocks of files and handle data read and write requests from clients. They also periodically report information about the data blocks they store and receive data operation instructions from the NameNode, including block replication, movement, and deletion.

### 2.3. Client

The client serves as the user interface of the file system, initiating file access requests and communicating with both the NameNode and DataNodes. Clients send file operation requests, such as creation, deletion, renaming, and locating, to the NameNode, which returns the results of file operations and related metadata information. Additionally, clients directly communicate with DataNodes to perform file data read and write operations. With file location information provided by the NameNode, clients access the file data stored on DataNodes directly.

### 2.4. Metadata Storage

In distributed file systems, persistent storage of the file system's metadata is essential for rapid recovery in the event of system restarts or failures. This metadata storage can take the form of a separate metadata server or a distributed metadata storage cluster. The NameNode assumes responsibility for managing and maintaining this metadata storage, ensuring consistency between it and the NameNode.

### 2.5. Fault Tolerance and Recovery Mechanisms:

High reliability is a crucial requirement for distributed file systems, necessitating fault tolerance and data recovery mechanisms. Data replication, redundancy, and recovery are commonly employed to achieve this goal. Data replication entails copying file data blocks among multiple DataNodes to improve data reliability and fault tolerance. In case of a DataNode failure, the system can retrieve data block backups from other nodes, enabling data recovery and automatic failover.

### 2.6. Load Balancing and Performance Optimization:

Distributed file systems must efficiently manage and schedule file access requests to fully utilize the storage and processing resources of each node. Load balancing mechanisms dynamically adjust the routing and distribution of file access requests based on node load conditions, optimizing system performance and maximizing resource utilization.

High-level distributed file system architecture aims to achieve high reliability, scalability, and performance optimization to meet the demands of large-scale data storage and access. Through well-designed and effectively managed components, distributed file systems can provide stable and reliable file storage services while meeting users' needs for data consistency and performance.

## 3. Overview of Classification Methods for Distributed File Systems

Distributed File Systems can be classified into multiple types based on different criteria. Firstly, they can be categorized into centralized and distributed metadata systems according to their architecture. Centralized distributed file systems store all metadata and control on a single server, while distributed metadata systems scatter metadata across multiple servers, with file data possibly distributed among multiple storage nodes. Secondly, they can be classified into strong consistency and eventual consistency based on data consistency models. Strong consistency distributed file systems ensure that reads of files always return the most recent write results, whereas eventual consistency distributed file systems may exhibit temporary inconsistencies after write operations, eventually converging to a consistent state. Lastly, they can be divided into block storage and object storage based on data access patterns. Block storage divides files into fixed-size blocks stored on storage nodes, while object storage stores files as objects, each with a unique identifier. Selecting the appropriate type of distributed file system should be based on the requirements and characteristics of the application scenarios. Here, we categorize distributed systems based on their architecture for discussion.

- Centralized Architecture:

  In a centralized architecture, there exists a central node or server responsible for managing the file system's metadata and controlling file access requests. All file access requests are routed through this central node, which is responsible for directing requests to the appropriate data nodes and coordinating data access operations. While this architecture is straightforward and easy to manage and implement, the central node may become a bottleneck, and there is a risk of a single point of failure[8].

- Distributed Architecture:

  A distributed architecture decentralizes the file system's metadata and control functions across multiple nodes, each capable of independently processing file access requests. This architecture eliminates the single point of failure associated with a central node and offers greater scalability and fault tolerance. Coordination mechanisms between nodes, such as distributed hash tables or consensus algorithms, maintain metadata consistency. Although distributed architectures typically require more complex coordination and communication mechanisms, they can provide higher performance and reliability.

- Hierarchical Architecture:

  A hierarchical architecture organizes the file system into multiple layers, each with distinct functions and responsibilities. These typically include storage, naming, and access layers. The storage layer handles actual data storage and management, the naming layer maintains the file system's namespace and metadata, and the access layer processes user file access requests and routes them to the appropriate storage nodes. Hierarchical architectures can enhance system scalability and performance and make the system easier to manage and maintain.

- P2P Architecture:

  In a peer-to-peer (P2P) architecture, all nodes are peers, sharing both the management and storage responsibilities of the file system. Each node acts as both a producer and consumer of data, enabling direct communication and data exchange between peers. P2P architectures offer high decentralization and fault tolerance, as there is no single point of failure. However, they can be more challenging to manage and maintain due to the participation of every node in system management and coordination.

These architectural classifications are not mutually exclusive, and practical distributed file systems may incorporate multiple architectural principles to meet various design requirements and scenario demands.

## 4. Challenges IN DISTRIBUTED FILE SYSTEM

When designing a distributed file system, various considerations need to be taken into account:

### 4.1. Performance

Performance is a key consideration when designing a distributed file system, involving response time, throughput, and concurrency. Optimizing performance involves implementing effective caching strategies, load balancing mechanisms, and data compression techniques to ensure the system can efficiently handle large amounts of data and user requests.

### 4.2. Reliability

Distributed file systems must have high reliability, ensuring data integrity and availability even in the event of node or network failures. Enhancing reliability involves techniques such as data backup, redundant storage, error detection, and correction to prevent data loss or corruption.

Nowadays, machine learning methods[9] can be employed for reliability analysis in distributed file systems. Simultaneously, by adopting optimization learning strategies, it becomes possible to more effectively analyze system data, extract valuable information, and subsequently optimize the system's operation. This can reduce the probability of failures, thereby enhancing the system's reliability.

*4.3. Transparency*

Distributed file systems should be transparent to users and applications, allowing them to access remote file systems as if they were accessing local file systems, without needing to be concerned about the underlying distributed architecture and implementation details. Transparency improves user experience and system usability. The different types of transparencies are as follows:

Table 1. Types of Transparencies [1].

| Transparency | Description |
| --- | --- |
| Access | Hides the differences in data representation and how a resource is accessed. |
| Location | Hides the physical location of a resource. |
| Migration | Hides the movement of a resource to another location. |
| Relocation | Hides the movement of a resource to another location while in use. |
| Replication | Hides the movement of a resource to another location while in use. |

*4.4. Security*

Ensuring the security of a distributed file system is crucial, including aspects such as data privacy protection, identity authentication, and access control. Implementing techniques such as encrypted transmission, access control lists, and authentication mechanisms can enhance system security, preventing unauthorized access and data leakage.

*4.5. Flexibility*

Distributed file systems should possess a degree of flexibility to adapt to different application scenarios and changing requirements. Flexibility includes supporting different storage media, file access methods, and data management strategies to meet diverse user needs.

*4.6. Scalability*

Distributed file systems should have good scalability, allowing them to scale horizontally or vertically as data volume and user demands increase. Employing distributed architectures and dynamic load balancing techniques can achieve scalability, ensuring the system can support large-scale data storage and access.

Table 2. Horizontal and Vertical Scaling of Distributed File Systems.

| Feature | Horizontal Scaling | Vertical Scaling |
| --- | --- | --- |
| Definition | Adding more nodes to increase capacity and performance | Enhancing the resources of a single node (e.g., CPU, memory) to improve performance and capacity. |
| Hardware Cost | Relatively low, as typically employs inexpensive hardware | Relatively high, as requires more powerful hardware to handle increased load. |
| Scalability | High, can gradually scale the system as needed | Limited, constrained by the physical architecture and performance of individual nodes. |
| Performance Impact | Increases linearly with the number of nodes added | Limited by the processing power and bandwidth of individual nodes |
| Fault Tolerance | High, replication and redundancy between nodes enhance system availability and fault tolerance | Low, as failure of a single node may result in system failure. |
| Use Cases | Suitable for scenarios requiring flexible capacity and performance scaling, such as cloud storage services | Suitable for scenarios where high availability and fault tolerance are required, and performance and capacity demands are not particularly high |

*4.7. Consistency*

Consistency in distributed file systems refers to the correctness and coherence of data across multiple nodes or replicas in the system. Achieving consistency is crucial for ensuring that all clients accessing the distributed file system observe a predictable and coherent view of the data, regardless of which node they access.

Consistency models define the rules and guarantees regarding the order and timing of read and write operations in a distributed system. There are several consistency models, each offering different trade-offs between consistency, availability, and partition tolerance:

- Strong Consistency[15]: Strongly consistent systems ensure that all read and write operations appear to occur instantly and atomically. Clients consistently observe the most recent data state, and updates are immediately visible across all nodes. Achieving strong consistency often involves coordination and synchronization mechanisms, which may impact system performance and availability.
- Eventual Consistency: Eventual consistency relaxes consistency requirements, allowing updates to propagate asynchronously across replicas. While temporary inconsistencies may exist between replicas, eventually, all replicas converge to the same state. Eventual consistency is often favored in distributed systems prioritizing low latency and high availability over strong consistency.
- Causal Consistency: Causal consistency preserves causal relationships between related operations. If operation A causally precedes operation B, then all replicas must observe operation A before operation B. Causal consistency strikes a balance between strong consistency and eventual consistency, ensuring correctly ordered causally related operations while permitting some degree of concurrency.
- Read-your-writes Consistency: This model guarantees that a client always sees the effects of its previous write operations when performing subsequent read operations. It is commonly used in distributed file systems to provide strong consistency for individual clients without imposing global synchronization overhead.
- Monotonic Consistency: Monotonic consistency ensures that once a client observes a particular data state, it will never observe a state preceding it. This prevents clients from regressing to previous states and is often used where data monotonicity is crucial, such as in distributed databases.
- Sequential Consistency: Sequential consistency guarantees that all operations appear to be executed sequentially, as if there were a single global timeline of operations. While providing strong consistency guarantees, enforcing sequential consistency in distributed systems can be challenging and often requires significant coordination overhead.

*4.8. Durability*

In the context of data storage, durability typically refers to the long-term preservation of data after it has been written to storage media, ensuring that data is not lost or corrupted due to system failures or power outages. Unlike reliability, durability focuses more on the security and protection of data to ensure that it can be recovered and accessed even in unforeseen circumstances.

Distributed file systems commonly employ replication to ensure data availability and durability in case of failures. Replication is often characterized by three parameters:
- n: the number of replicas
- r: the number of replicas read by a reader
- w: the number of replicas to which a writer synchronously writes, requiring acknowledgment before confirming to the client.

When r + w > n, strong consistency[15] is achieved, ensuring that readers see all writes, though they may need to resolve any conflicting updates not yet addressed by the file system. To maintain strong consistency, a typical method involves designating a primary node for each data item, which must be one of the w nodes acknowledging the write synchronously. This allows readers to contact the primary node for strong consistency or any node if reading slightly stale data is acceptable.

Considering the above factors comprehensively can help designers create efficient, stable, and secure distributed file systems to meet the needs of different users and applications.

**5. Common distributed file systems**

*5.1. Network File System*
NFS [10] was initially developed by a group of engineers at Sun Microsystems to provide file sharing capabilities for the company's UNIX systems. The first version of NFS was released in 1985, using the UDP protocol for communication. NFS v3, released in 1995, marked a significant improvement, introducing many performance and feature enhancements, including support for the TCP protocol, improved caching mechanisms, and better file locking support. In 2000, NFS v4 introduced many new features, such as enhanced security, file system abstraction, and stronger locking mechanisms, along with support for IPv6.

NFS adopts a client-server architecture where each file server offers a standardized view of its local file system. Clients access the server transparently through an interface similar to the local file system interface. Client-side caching may be used to save time and network traffic. The server defines and executes all file operations.

The Virtual File System (VFS) acts as an interface between the operating system's system call layer and all file systems on a node. The user interface to NFS is the same as the interface to local file systems. Calls are forwarded to the VFS layer, which then passes them to either the local file system or the NFS client.

NFSv4 is stateful, and clients communicate with servers using RPCs. NFS employs various mechanisms to ensure data synchronization and consistency, thereby improving system reliability. Firstly, file system operations are synchronous between clients and servers, ensuring that each operation is correctly executed and acknowledged to prevent data inconsistency. Secondly, NFS provides file-level locking mechanisms, allowing clients to lock files to prevent data corruption from concurrent write operations. Additionally, NFS uses caching mechanisms between clients and servers to enhance performance while ensuring data consistency, such as through write acknowledgment mechanisms and management of data cache consistency. Despite NFS's efforts to ensure data synchronization and consistency, data inconsistency may still occur in cases of network or server failures, necessitating careful handling of data synchronization and consistency issues when using NFS. NFS employs various mechanisms to enhance system fault tolerance, including connection retries, status monitoring, caching mechanisms, and retry mechanisms, to ensure effective handling and recovery in the event of failures or connection interruptions.

*5.2. Andrew File System*
AFS (Andrew File System) is a distributed file system developed by Carnegie Mellon University, initially in the early 1980s. Originally aimed at addressing file sharing and management issues faced by Carnegie Mellon University, AFS introduced a global namespace, allowing users to access files across the network in a uniform manner [12]. It also provided high scalability and availability by distributing copies of file data among multiple servers and caching data among clients to improve performance. Over time, AFS has become a standard distributed file system in research and academic circles, with some application in the commercial sector as well.

In AFS, a cell serves as a fundamental organizational unit, managing a group of server and client machines. Each AFS cell is administered independently by a designated administrator. Typically composed of a set of interconnected servers and clients, an AFS cell forms a logical file system unit. Servers within an AFS cell are responsible for storing and managing file data, while clients access and utilize this data. The creation and management of cells can be adjusted and configured according to specific needs to meet the requirements of different organizations or environments. Each cell in AFS can be viewed as an independent file system instance with its own namespace and permission settings, and each cell is identified by a unique name, which locates it within the entire AFS environment. Servers and clients can only belong to a single cell at a time, although users can have accounts in multiple cells. From the user's perspective, the AFS hierarchy is accessed through the path /afs/[cell name] using standard file system operations. Since AFS is location-transparent, the path to a file in the tree remains the same regardless of the client's location.

Servers and clients in AFS are stateless and utilize a communication mechanism called RPC2 (Remote Procedure Call Version 2). RPC2 is a custom communication protocol designed to provide efficient remote procedure call functionality, enabling communication between clients and servers and supporting data transfer and command execution. RPC2 uses UDP or TCP as the transport protocol and provides a reliable communication mechanism to ensure data integrity and reliability.

AFS provides file-level locking mechanisms, allowing clients to lock files to prevent data corruption caused by concurrent write operations. This ensures that while one client is modifying a file, other clients cannot simultaneously modify the same file, maintaining data consistency. File system operations between clients and servers are synchronous, meaning each operation is confirmed and executed between the client and server to ensure data consistency and correctness, preventing data inconsistencies. AFS utilizes caching mechanisms between clients and servers to improve performance while also serving as a fault tolerance mechanism. Clients cache recently accessed file and directory information to reduce reliance on servers. Even in the event of a lost connection with the server, clients can still access data from their local cache, ensuring a certain level of availability.

*5.3. Google File System*

The Google File System (GFS) is a distributed file system developed by Google, first publicly disclosed in 2003. The design of GFS was inspired by Google's deep understanding of the need for large-scale data storage [12]. At that time, Google needed to handle massive datasets and provide a highly reliable and high-performance storage solution to support the operation of its search engine and other services. To address these challenges, Google developed GFS as the foundational infrastructure for its internal storage system.

The design goals of GFS are to provide high fault tolerance, scalability, and performance. GFS adopts a master-slave architecture, where a GFS cluster consists of a master node (Master) and multiple chunk servers (Chunkserver). The master node is the central node of the GFS cluster, responsible for managing file system metadata, coordinating data storage and access operations, monitoring the health status of the cluster, and more. Chunk servers are storage nodes in the GFS cluster, responsible for storing actual data blocks and handling read and write requests from clients. In GFS, files are divided into fixed-size data blocks (Chunks), typically 64MB in size, and redundant backup strategies are used to ensure data safety. Each data block is replicated on multiple chunk servers, usually with 3 replicas. The master node is responsible for selecting the primary replica of the data block and coordinating replication and consistency among replicas.

GFS servers are stateful, as they need to maintain all the state information of client requests. GFS RPC is a custom remote procedure call protocol defined in GFS, based on TCP, used for communication between clients, master nodes, and chunk servers. Clients send requests to the master node via GFS RPC to obtain file metadata information, file location information, etc. The master node is responsible for responding to and processing these requests, returning the corresponding metadata information or executing the corresponding management operations. Clients send read and write requests to chunk servers, which handle these requests and read from or write to the corresponding data blocks.

GFS uses various concurrency control and locking mechanisms to ensure consistency and correctness when multiple clients access the same file concurrently. For example, GFS provides file-level locking mechanisms, allowing clients to lock files to prevent data corruption caused by concurrent write operations. When multiple clients write to the same data block simultaneously, GFS determines the order of write operations based on version numbers and ensures that data updates are performed in the correct order. When clients write data to chunk servers, the servers must send write acknowledgment messages to the clients, and clients only consider write operations successful after receiving a sufficient number of acknowledgment messages. GFS clusters have high fault tolerance, automatically detecting and recovering replicas of data blocks. The master node uses persistent storage to store the file system's metadata, preventing data loss caused by master node failures.

*5.4. Hadoop Distributed File System*

Apache Hadoop Distributed File System (HDFS) is one of the core components of the Apache Hadoop project, developed by the Apache Software Foundation. Initially released in 2006 as an open-source implementation of the Google File System, HDFS became part of the Apache Hadoop project in 2008 [13]. Inspired by the Google File System, its design aims to provide a reliable and efficient distributed file storage solution for large-scale data processing, with goals including high reliability, high throughput, and fault tolerance. As big data technology has evolved and application scenarios have expanded, HDFS has gradually become one of the industry-standard distributed file systems, widely used in various large-scale data processing and analytics tasks.

HDFS adopts a master/slave architecture. The NameNode, a core component of HDFS, is responsible for storing the file system's metadata, including the file tree, namespace, file attributes, and block lists. It maintains the namespace, directory structure, and mapping of files to blocks for the entire file system. However, due to being a single point of failure, it requires backup or adoption of high-availability solutions such as Secondary NameNode or High Availability (HA). The DataNode is a node that stores actual data blocks, responsible for storing and retrieving file data blocks. Each DataNode manages the storage of the local file system and periodically reports block information to the NameNode. The client is an external entity that interacts with HDFS, responsible for sending file operation requests such as reading, writing, and deleting files to HDFS. Clients execute these operations by communicating with the NameNode and DataNode and accessing the file system through the HDFS API. HDFS supports parallel processing of large-scale data, with data streams parallelly read and written through multiple DataNodes to improve throughput and performance.

HDFS servers are Stateful. The communication protocol used in HDFS is mainly based on a custom Remote Procedure Call (RPC) protocol over TCP/IP. This RPC protocol is known as the Hadoop RPC protocol, allowing communication between different components, such as between clients and NameNodes, DataNodes, and between DataNodes. HDFS ensures data reliability and fault tolerance through data replication, where each data block usually has multiple replicas stored on different DataNodes. The NameNode is responsible for managing replica location information, block replication policies, and handling replica replication and movement.

When writing data, HDFS first writes the data to the local file system of the DataNode and then sends metadata information such as block location and checksum to the NameNode for recording. Only after the NameNode confirms receiving this information does it return a successful write message to the client. Additionally, HDFS provides a file-level locking mechanism, allowing clients to lock files to prevent data corruption from concurrent write operations. When multiple clients simultaneously modify the same file, HDFS determines the order of write operations based on version numbers and ensures data updates are performed in the correct order.

*5.5. Ceph File System[14]*

CephFS is an open-source distributed file system that was initially developed by Sage Weil in 2006 and first publicly released in 2007. Initially, CephFS was developed as part of the Ceph storage system, but over time, it evolved into a standalone file system project and gained widespread attention and usage in the open-source community.

The metadata server is one of the key components of CephFS, responsible for storing and managing the file system's metadata, including directory structure, file attributes, and access permissions. In CephFS, the operation of metadata servers is similar to that of traditional distributed file systems' metadata servers but with higher scalability and performance. These metadata servers are typically deployed in a cluster, consisting of multiple active and standby servers to ensure high availability and fault tolerance.

Object storage devices are also crucial components of the Ceph storage cluster, responsible for storing file data blocks and other data required by metadata servers. OSDs handle read, write, and replication operations and ensure the reliability and consistency of data. File data blocks in CephFS are stored in OSDs and managed and accessed through metadata servers. CephFS also supports file-level locking mechanisms, allowing clients to lock files to prevent data corruption from concurrent write operations.

The communication mechanism of CephFS is based on RADOS (Reliable Autonomic Distributed Object Store), which is a core component of the Ceph storage system. RADOS provides underlying storage support for CephFS and offers distributed object storage functionality. It manages object storage devices (OSDs) and provides services for storing, retrieving, and replicating data. CRUSH (Controlled Replication Under Scalable Hashing) is an algorithm used in the Ceph storage system for data distribution and replication. This algorithm dynamically determines the location of data in the storage cluster based on the data's hash value and the topology of storage devices, ensuring balanced distribution and high availability of data. Moreover, it can further facilitate the improvement of algorithmic implementation for machine learning [16-19].

## 6. CONCLUSION

Distributed File System is an advanced solution designed to efficiently store and manage file data. By dispersing file data across multiple interconnected computer nodes, DFS achieves the goal of decentralized file storage management. The core objective of DFS is to provide a flexible, scalable, and resilient file storage framework, enabling users to seamlessly access and manipulate files distributed across different nodes. The design of DFS involves several key components and design considerations, such as NameNode, DataNodes, metadata storage, fault tolerance and recovery mechanisms, as well as load balancing and performance optimization. DFS can be classified based on different criteria, such as architecture, consistency models, and data access patterns. When designing DFS, it is essential to consider various factors such as performance, reliability, transparency, security, flexibility, scalability, and consistency. Common distributed file systems include NFS, AFS, GFS, HDFS, and CephFS, each with specific design goals and use cases.